\documentclass[prb,twocolumn,superscriptaddress,showpacs,floatfix]{revtex4}
\makeatletter\def\Dated@name{}\makeatother

\preprint{EKM-TP3/10-06}

\newcommand{\picturewidth}{\hsize}

\usepackage{amsmath,amssymb,amsfonts}
\usepackage{dcolumn}
\usepackage{bm}
\usepackage{graphicx,color}

\newcommand{\bmSigma}{{\bm{\Sigma}}}
\newcommand{\bmG}{{\bm{G}}}

\newcommand{\Tr}{{\text{Tr}}}
\newcommand{\F}{{\text{F}}}
\newcommand{\bra}{\langle}
\newcommand{\ket}{\rangle}

\begin{document}

  \title{Phase separation in the particle-hole asymmetric Hubbard model}

  \author{Martin Eckstein}
 
  \author{Marcus Kollar}
  \affiliation{Theoretical Physics III, Center for
    Electronic Correlations and Magnetism,\\ Institute for Physics,
    University of Augsburg, 86135 Augsburg, Germany}
 
  \author{Michael Potthoff}
  \affiliation{Institute for Theoretical Physics and Astrophysics,
  University of W\"{u}rzburg,\\ Am Hubland, 97074 W\"{u}rzburg, Germany}
 
  \author{Dieter Vollhardt}
  \affiliation{Theoretical Physics III, Center for
    Electronic Correlations and Magnetism,\\ Institute for Physics,
    University of Augsburg, 86135 Augsburg, Germany}
 
  \date{October 30, 2006}

  \begin{abstract}
    The paramagnetic phase diagram of the Hubbard model with
    nearest-neighbor (NN) and next-nearest-neighbor (NNN) hopping on
    the Bethe lattice is computed at half-filling and in the weakly
    doped regime using the self-energy functional approach for
    dynamical mean-field theory.  NNN hopping breaks the particle-hole
    symmetry and leads to a strong asymmetry of the electron-doped and
    hole-doped regimes. Phase separation occurs at and near
    half-filling, and the critical temperature of the Mott transition
    is strongly suppressed.
  \end{abstract}

  \pacs{71.27.+a, 71.30.+h\\[0pt]}

  \maketitle
  
  \section{introduction}
  \label{section:intro}

  One of the major goals of condensed-matter physics during the past
  decades has been to understand the role of electronic correlations
  in solids.  At the heart of this field lies the metal-insulator
  transition (MIT) driven by the electronic interaction, the Mott
  transition.\cite{mott} It occurs in many materials,\cite{imada-1998}
  in particular transition-metal compounds such as $\rm V_2O_3$.  The
  underlying physical mechanism for this transition is captured by the
  single-band Hubbard model,\cite{hubbard-1963}
  \begin{equation}
    H=
    \sum_{ij\sigma} t_{ij}
    c_{i\sigma}^\dagger c_{j\sigma}
    + U\sum_{i} n_{i\uparrow}n_{i\downarrow}
    -\mu\sum_{i\sigma}n_{i\sigma}\,.
    \label{eq:hubbard}
  \end{equation}
  Here $c_{i\sigma}^\dagger$ are creation operators for an electron at
  site $i$ with spin $\sigma$, $U$ is the local Coulomb repulsion, and
  $t_{ij}$ are the hopping amplitudes, e.g.,
  $t_1$ for hopping between nearest neighbors (NN) and
  $t_2$ for hopping between next-nearest neighbors (NNN).
  In the Hubbard model, the Mott
  transition can be driven either by an increase of the interaction
  $U$ (at half-filling), or, for large enough $U$, by a change of the
  particle density towards one electron per site.  The two cases are referred
  to as bandwidth-controlled Mott transition and filling-controlled
  Mott transition, respectively.\cite{imada-1998}
  
  Much progress in understanding electronic correlation effects in
  general, and the Mott transition in particular, was made by
  considering the limit of infinite dimensions.\cite{metzner-1989b}
  Dynamical mean-field theory (DMFT) \cite{georges-1996,kotliar-2004}
  can then be used as a unified framework to study metallic,
  insulating, and magnetically ordered phases. Both the
  bandwidth-controlled
  MIT\cite{jarrell-1992,georges-1993,rozenberg-1994,georges-1996,
    bulla-1999,rozenberg-1999,bulla-2001,tong-2001,joo-2001,bluemer-phd}
  and the filling-controlled
  MIT\cite{georges-1996,fisher-1995,kajueter-1996,ono-2001,kotliar-2002,garcia-2006,werner-2006}
  were studied within DMFT; for $T>0$ the transition is first-order.
  However, in the case of a bipartite lattice and if $t_{ij}$ is
  restricted to NN hopping, i.e., in the particle-hole
  symmetric case, the Mott transition at half-filling quite generally
  does not occur since it is preempted by the intrinsic weak-coupling
  instability towards antiferromagnetic (AF) long-range order.  To
  reduce or even avoid this ``parasitic'' AF phase one may consider
  non-bipartite lattices\cite{ohashi-2006,aryanpour-2006} or
  alternatively introduce NNN hopping.
  The latter introduces a competition which frustrates the
  antiferromagnetic order.  It was proposed that the phase diagram of
  the single-band Hubbard model with NN and NNN hopping then
  qualitatively resembles that of various correlated-electron
  materials and thus appears to be the minimal model for such
  materials.\cite{rozenberg-1995,kotliar-2004}
  
  In addition to the frustration of the antiferromagnetic phase, NNN
  hopping generally makes the Hubbard model particle-hole asymmetric,
  even at half-filling.  This asymmetry, which is a generic property
  of real materials,\cite{hirsch-2002,hirsch-2005} will also have an
  effect on the paramagnetic phase.  However, previous DMFT
  investigations considered a special type of hopping (``random $t_2$
  hopping'') in which the model remains particle-hole symmetric even
  for $t_2\ne0$, and the paramagnetic phase is not changed at
  all.\cite{georges-1996,rozenberg-1995,rozenberg-1994,chitra-1999,zitzler-2004}
  By contrast, the effect of non-random NNN hopping becomes evident
  already in the non-interacting system through an asymmetric density
  of states (DOS), as derived recently by a new theoretical technique
  which is able to treat arbitrary hopping on the Bethe
  lattice.\cite{eckstein-2005,kollar-2005} In the present paper we
  proceed to discuss the effects of NNN hopping $t_2$ on
  the phase diagram for finite $U$ and $T$.
  
  As we will show, a striking consequence of NNN hopping is the
  occurrence of phase separation between metallic and insulating
  phases even {\em at half-filling}. Indeed, in the paramagnetic phase of the
  Hubbard model, phase separation is known to lead to hysteresis in the density
  $n(\mu)$ at the Mott transition.\cite{kotliar-2002,macridin-2006}
  For the Hubbard model on a square lattice
  ($d=2$), which is closely related to high-$T_c$ superconductors,
  phase separation was found within the dynamical cluster
  approximation (DCA) for $t_2>0$ and $n<1$,\cite{macridin-2006}
  whereas for $t_2=0$ finite-size Monte Carlo
  calculations\cite{moreo-1991,becca-2000} provide evidence for a
  homogeneous state.  On the other hand, in infinite spatial dimensions DMFT
  predicts phase separation already for $t_2=0$ and
  $n\ne1$,\cite{kotliar-2002,werner-2006} but not at half-filling.
  We find that this changes for $t_1$-$t_2$ hopping.
  
  For our investigation of the effect of $t_2$ hopping on the
  paramagnetic phase diagram we use the recently developed self-energy
  functional approach (SFA).\cite{potthoff-2003a} The SFA is a
  variational method based on an exact variational principle for the
  grand potential.\cite{luttinger-1960} In this scheme, the
  self-energy of a finite reference system is taken as variational
  ansatz for the self-energy of the lattice system.  The method is
  very general and can be applied to finite-dimensional
  systems\cite{potthoff-2003c} and for various types of
  interactions.\cite{tong-2005} Within DMFT the exact result would be
  recovered for a single-impurity Anderson model (SIAM) with
  infinitely many bath sites as reference system,\cite{potthoff-2003b}
  while in practice only a SIAM with a finite number $n_s$ of bath
  sites can be investigated. However, the results converge quickly
  with increasing $n_s$.\cite{pozgajcic-2004} The simplest impurity
  model considers only a single bath site.  Surprisingly, this
  reference system is already enough to reproduce previous results for
  $t_2=0$ for the Mott transition qualitatively and, sometimes, even
  quantitatively.\cite{potthoff-2003b} Therefore we will use this
  ``two-site approximation'' (one correlated site, one bath site) also
  to investigate the case $t_2\ne0$.
  
  The paper is organized as follows: In Sec.~\ref{section:methods} we
  give a short description of the SFA and of our method to treat NNN
  hopping on the Bethe lattice. In Sec.~\ref{section:NN} NN
  hopping is studied at and away from half-filling.  We calculate the
  paramagnetic phase diagram as a function of interaction,
  temperature, and either chemical potential or density.  The phase
  diagrams for NN and NNN hopping are discussed in detail in
  Sec.~\ref{section:t1t2}.

  \section{Models and Methods}
  \label{section:methods}
  
  \subsection{Self-energy functional approach}

  The self-energy functional approach (SFA) is based on the
  variational principle of Luttinger and Ward.\cite{luttinger-1960}
  When formulated in terms of the self-energy $\bmSigma$, the basic
  quantity in this context is the grand-potential functional
  $\hat{\Omega}[\bmSigma]$, which becomes stationary at the physical
  self-energy.\cite{potthoff-2003a} It can be written as
  \begin{equation}
    \hat{\Omega}[\bmSigma] =
    \hat{\F}[\bmSigma]
    +\Tr\log[-(\bmG_0^{-1}-\bmSigma)^{-1}]\,,
    \label{eq:omega(sigma)}
  \end{equation}
  where $\hat{\F}[\bmSigma]$ contains all information about the
  interaction, while the noninteracting Green function $\bmG_0$ is
  considered fixed. (Quantities written in boldface are matrices in
  single-particle indices and Matsubara frequencies, and $\Tr$ denotes
  the corresponding trace.)  The functional form of $\hat{\F}[\bmSigma]$
  is generally unknown, apart from the fact that it depends only on
  the interaction part in the Hamiltonian, and not on the kinetic
  energy.  This universality follows directly from various ways in
  which $\hat{\F}[\bmSigma]$ or its Legendre transform, the
  Luttinger-Ward functional, can be constructed.\cite{potthoff-2004}
  It allows one to calculate $\hat{\Omega}[\bmSigma]$ exactly for
  certain variational self-energies, namely for all $\bmSigma$ that
  can be considered as the exact self-energy of a reference system
  with the {\em same} interaction. The self-energy, $\bmSigma=\bmSigma(y_i)$ 
  is varied by varying single-particle parameters $\{y_i\}$ of that 
  reference system.
  In general $\bmSigma$ can only be calculated
  for small systems, by means of exact diagonalization. 
  Comparing with the self-energy functional of the reference system
  and exploiting the universality of $\hat{\F}[\bmSigma]$, 
  Eq.~(\ref{eq:omega(sigma)}) can be written as:
  \begin{eqnarray}
    \hat{\Omega} [\bmSigma] & = & \Omega'
    -  \Tr \log [-(\bmG_0'^{-1}-\bmSigma)^{-1}]
    \nonumber\\
    & & +\,  \Tr \log [-(\bmG_0^{-1}-\bmSigma)^{-1}],
    \label{eq:sfa}
  \end{eqnarray}
  where $\Omega'$ and $\bmG_0'$ correspond to the exact grand
  potential and the noninteracting Green function of the reference
  system, respectively. The best approximation for the self-energy is
  then determined from $\partial\Omega[\bmSigma(y_i)]/\partial y_i=0$.
  
  As noted in the Introduction, the reference system employed in our
  investigation is a two-site impurity model,\cite{potthoff-2003b}
  \begin{equation}
    H_{ref}=U n_{1\uparrow}n_{1\downarrow} +\sum_{i=1,2 \sigma} \epsilon_{i} n_{i\sigma}
    +\sum_{\sigma}V(c_{1\sigma}^\dagger c_{2\sigma}+\rm{h.c.})\,.
  \end{equation}
  Here index $1$ refers to the correlated impurity and index $2$ to
  the bath site.  The on-site energies $\epsilon_{1,2}$ and the
  hybridization $V$ are taken as variational parameters.  As we are
  interested only in the paramagnetic phase and thus need to consider
  only spin-independent self-energies, we can take $\epsilon_i$ and
  $V$ to be spin-independent as well.
  
  Once a stationary point of $\Omega$ is found, it is tracked in the
  space of variational parameters $\{y_i\}$ as a function of the
  external physical parameters (interaction $U$, temperature $T$, and
  chemical potential $\mu$) using local algorithms for the solution of
  the equations $\partial\Omega/\partial y_i=0$.  High accuracy is
  achieved by calculating derivatives of $\partial\Omega/\partial y_i$
  analytically.\cite{pozgajcic-2004} However, close to a first-order
  transition it is inconvenient to parametrize the solution by the
  physical parameters, because in this parametrization several
  solutions coexist which correspond to various metastable phases.  To
  avoid repeated switching between these solutions we use an algorithm
  that treats the physical and variational parameters on equal
  footing.

  \subsection{NN and NNN hopping on the Bethe lattice}

  The properties of the lattice enter Eq.~(\ref{eq:sfa}) only via the
  term $\Tr\log[-(\bmG_0^{-1}-\bmSigma)^{-1}] \equiv
  \hat{\Omega}_{\text{latt}}[\bmSigma]$.  In this subsection we show how to
  evaluate $\hat{\Omega}_{\text{latt}}$ for arbitrary hopping on the Bethe
  lattice, in particular for $t_1$-$t_2$ hopping.  We restrict the
  analysis to the paramagnetic phase with a spin- and site-independent
  self-energy $\bmSigma\equiv\Sigma(i\omega_n)$.  Generally the trace
  in $\hat{\Omega}_{\text{latt}}$ includes a sum over Matsubara frequencies and
  over all eigenstates $|k\ket$ of the matrix
  $\bmG_0^{-1}-\bmSigma=i\omega_n+\mu-\Sigma(i\omega_n)-{\bm t}$,
  where ${\bm t}$ is the matrix of hopping amplitudes.  The sum over
  $|k\ket$ is then expressed in terms of a sum over lattice sites $i$
  and an integral over the local density of states,
  $\rho_i(\epsilon)= \sum_k \delta(\epsilon-\epsilon_k)|\bra i | k
  \ket|^2$, and the sum over Matsubara frequencies is transformed into
  an integral over real frequencies as done in
  Ref.~\onlinecite{potthoff-2003b}.  This leads to the expression
  \begin{equation}
    \hat{\Omega}_{\text{latt}} =
    2\sum_{i} \int\limits_{-\infty}^{\infty} d\omega f(\omega)
    \int\limits_{-\infty}^{\infty} d\epsilon \,
    \rho_i(\epsilon) \Theta[\omega+\mu-\Sigma(\omega)-\epsilon] \, ,
    \label{eq:olatt1}
  \end{equation}
  where $f(\omega)$ is the Fermi function and the factor $2$ accounts
  for spin degeneracy.  For the Bethe lattice with arbitrary
  coordination number $Z$ there exists a general method to calculate
  the DOS for any given hopping
  Hamiltonian.\cite{eckstein-2005,kollar-2005} One first introduces
  hopping Hamiltonians $H_d=\sum_{d_{ij}=d} | i \ket \bra j |$ that
  describe hopping only between $d$th nearest neighbors. In the limit
  of infinite connectivity $K=Z-1$ these Hamiltonians must be
  scaled\cite{metzner-1989b} according to $\tilde H_d = H_d/ K^{d/2}$,
  where $\tilde H_d$ retains a nontrivial spectrum for $K\to\infty$.
  In particular, the spectrum of the NN hopping Hamiltonian $\tilde
  H_1$ has the well-known semi-elliptical form $\rho_0(\lambda) =
  \sqrt{4-\lambda^2}/(2\pi)$ in this limit.  One may then use a
  remarkable relation valid for the Bethe lattice, which makes use of
  its special topological properties.\cite{eckstein-2005} 
  Namely, every hopping Hamiltonian can
  be written as a function of $\tilde H_1$, i.e., for an arbitrary set
  of hopping amplitudes $\{t_d^*\}$ one has
  \begin{equation}
    \sum_d t_d^* \tilde H_d = \epsilon(\tilde H_1)\,.
    \label{eq:dispersion}
  \end{equation}
  Analytical expressions for $\epsilon(\lambda)$ were derived in
  Ref.~\onlinecite{eckstein-2005}.  In analogy to crystal lattices,
  where any translationally invariant Hamiltonian is a function of
  momentum, $\epsilon(\lambda)$ plays the role of a dispersion
  relation, and $\lambda\in[-2,2]$ runs over the spectrum of $\tilde
  H_1$. In particular, the dispersion for $t_1$-$t_2$ hopping is given
  by \cite{eckstein-2005}
  \begin{equation}
    \epsilon(\lambda) = t_2^* \lambda^2 + t_1^* \lambda -t_2^*.
    \label{eq:t1t2-dispersion}
  \end{equation}
  The DOS can then be obtained in a straightforward way from the known
  DOS for $\tilde H_1$ by a change of variables.  The lattice
  contribution for arbitrary range hopping then reduces to
  \begin{equation}
    \hat{\Omega}_{\text{latt}}=
    \int\limits_{-\infty}^{\infty} d\omega f(\omega)
    \int\limits_{-2}^{2} d\lambda \frac{\sqrt{4-\lambda^2}}{\pi}
    \Theta[\omega + \mu -\Sigma(\omega) - \epsilon_{t_d^*}(\lambda)]\,.
    \label{eq:olatt}
  \end{equation}
  This expression is suitable for numerical evaluation.  For
  $t_1$-$t_2$ hopping the inner integral can be evaluated analytically
  by solving a quadratic equation. This allows us to obtain
  $\hat{\Omega}_{\text{latt}}$ with great precision, which is necessary to
  determine the stationary points of $\hat{\Omega}$ reliably.

  We conclude this section with a general remark about the sign of the
  hopping amplitudes $t_1$ and $t_2$. For any bipartite lattice the
  unitary transformation $c_{i\sigma}\to(-1)^i c_{i\sigma}$, where
  $(-1)^i$ is alternating on the two sublattices, changes the sign of
  all hopping amplitudes between the two sublattices. Thus we can
  assume $t_1>0$ without loss of generality.  On the other hand, it
  follows from the particle-hole transformation $c_{i\sigma}\to(-1)^i
  c_{i\sigma}^\dagger$ that $t_2\to-t_2$ merely interchanges
  electron-doped ($n>1$) and hole-doped ($n<1$) regimes. Therefore we
  can also assume $t_2>0$.
  
  \section{NN hopping}
  \label{section:NN}

  The paramagnetic phase-diagram of the Hubbard model with only NN
  hopping ($t_2=0$) has been studied intensively using DMFT, both for
  half-filling\cite{georges-1996,bulla-2001,tong-2001,joo-2001} and
  for finite
  doping.\cite{georges-1996,fisher-1995,kajueter-1996,kotliar-2002,garcia-2006,werner-2006}
  In this section we use the two-site approximation to investigate the
  full parameter space spanned by $U$, $T$, and $\mu$, and compare
  with previous results where available.
  \begin{figure}
    \centerline{\includegraphics[clip,width=\picturewidth,angle=0]{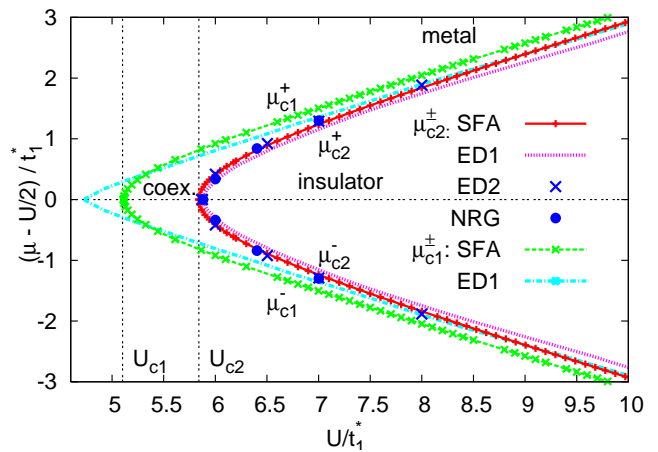}}
    \caption{$(U,\mu)$ phase diagram at $T=0$ for NN hopping.
      Results from two-site SFA (this work) are compared with ED (ED1:
      8 sites\cite{kajueter-1996}, ED2: extrapolation from 11
      sites\cite{ono-2001}) and NRG.\cite{ono-2001} Due to
      particle-hole symmetry the phase diagram is symmetric with
      respect to $\mu=U/2$.}
    \label{fig:NN-T=0}
  \end{figure}

  At $T=0$ the phase diagram has a simple structure
  (Fig.~\ref{fig:NN-T=0}).  At half-filling (where $\mu=U/2$) the
  system is insulating for $U>U_{c2}$, where $U_{c2} \approx 5.84
  t_1^*$ for the semi-elliptic DOS.\cite{bulla-2001} For any other
  filling the ground state is metallic.  In the metal the charge
  compressibility $\chi_{\mu}= \partial n/\partial\mu$ remains finite
  even when the insulator is approached from larger and lower filling,
  but the chemical potential changes discontinuously from
  $\mu=\mu_{c2}^-$ to $\mu=\mu_{c2}^+$ as the filling goes from
  $n=1^-$ to $n=1^+$. For $\mu_{c2}^-<\mu<\mu_{c2}^+$ the system is
  insulating and half-filled (Fig.~\ref{fig:NN-T=0}).  In a region
  $\mu_{c1}^-<\mu<\mu_{c2}^-$ and $\mu_{c1}^+>\mu>\mu_{c2}^+$ both
  metallic and insulating DMFT solutions exist, with the metallic
  phase being the thermodynamically stable one.  The appearance of the
  metal at $\mu=\mu_{c2}^\pm$ is related to the development of states
  inside the Mott gap, whereas the breakdown of the insulator at
  $\mu=\mu_{c1}^\pm$ occurs when $\mu$ reaches the edge of the
  gap.\cite{fisher-1995} In Fig.~\ref{fig:NN-T=0} we display the phase
  diagram obtained with the two-site approximation and compare with
  exact diagonalization (ED).\cite{kajueter-1996,ono-2001} The
  agreement is remarkably good, in particular concerning the value of
  $\mu_{c2}^\pm$.  Note that both methods are essentially based on an
  approximation for the self-energy by a rational function, but the
  two-site approximation uses a function of much simpler structure,
  i.e., with fewer poles.

\begin{figure}[t]
    \centerline{\includegraphics[clip,width=\picturewidth,angle=0,bbllx=87,bblly=77,bburx=367,bbury=277]{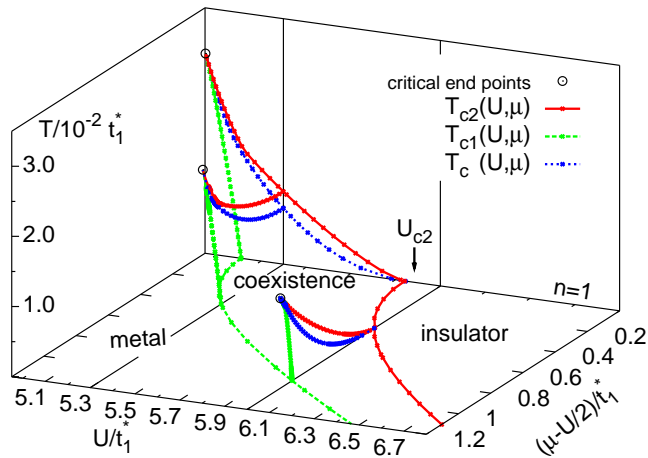}}
    \caption{($U,T,\mu$) phase diagram for $t_2=0$
      in two-site SFA. The plot
      is constructed from the ($U,\mu$) phase
      diagram at $T=0$ (Fig.\ \ref{fig:NN-T=0}),
      the ($T,U$) phase diagram at half-filling ($\mu=U/2$)
      as in Ref.~\onlinecite{potthoff-2003b},
      and two ($T,\mu$) sections
      at fixed interaction $U=5.3\,t_1^*$ and $U=6\,t_1^*$.
      Due to particle-hole symmetry the phase 
      diagram is symmetric with respect
      to $\mu=U/2$. Only the electron-doped region
      ($\mu>U/2$) is displayed.}
    \label{fig:NN-UTmu}
  \end{figure}

  The extension of these results to $T>0$ is best described in the
  grand canonical ensemble.  In Fig.~\ref{fig:NN-UTmu} four
  representative planes in the ($U$,$T$,$\mu$) phase diagram are
  shown: The ($T,U$) plane at half-filling\cite{potthoff-2003b}
  ($\mu=U/2$), the ($U,\mu$) plane at $T=0$ discussed in the last
  paragraph, and two ($T$,$\mu$) planes at fixed interaction. The
  latter interactions are $U=5.3 t_1^*$ and $U=6 t_1^*$, which are
  slightly smaller and larger than $U_{c2}$, respectively.
  Our results show that the metallic and insulating regions are
  separated by a first-order transition {\em surface} $T_c(U,\mu)$
  which coincides with the transition line $T_c(U)$ at $\mu=U/2$ and
  with $\mu_{c2}^\pm$ at $T=0$.  The coexistence region associated
  with this transition lies between two surfaces $T_{c1}(U,\mu)$ and
  $T_{c2}(U,\mu)$ where the metastable insulating and metallic
  solutions disappear. They intersect the $T=0$ plane in the lines
  $\mu_{c1}^\pm(U)$ and $\mu_{c2}^\pm(U)$ and meet in a line of
  second-order critical endpoints.

\begin{figure}[t]
    \centerline{\includegraphics[clip,width=\picturewidth,angle=0]{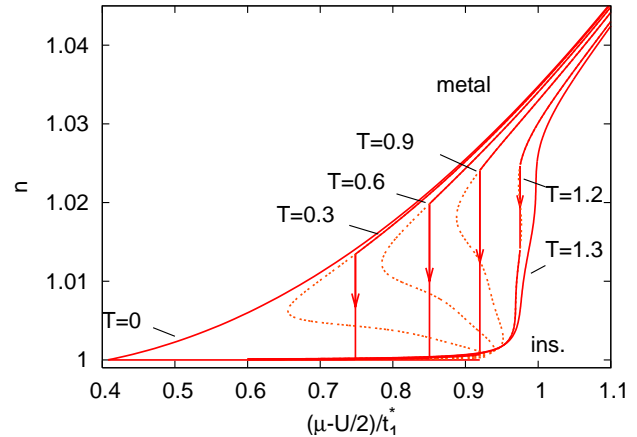}}
    \caption{Density as a function of the chemical potential for
      $U=6t_1^*$ and $t_2=0$.  The temperature values are given in
      units of $10^{-2}\,t_1^*$.  Solid lines correspond to the stable
      phases, while dotted lines correspond to the metastable metallic ($\mu<\mu_{c}$)
      and insulating ($\mu>\mu_{c}$) phase, as well as to a third
      stationary point of the grand potential functional, which does
      not correspond to any physical phase because $\chi_\mu<0$ (in
      analogy to the $p$-$V$ diagram of the van-der-Waals gas).  The
      discontinuity of $n$ at the transition from the larger value in
      the metal to a value close to $n=1$ in the insulator is
      indicated by an arrow.}
    \label{fig:filling}
  \end{figure}

  To detect phase separation we calculate the density $n(\mu)$ for
  given $U$ and various $T>0$ as shown in Fig.~\ref{fig:filling}.  For
  the first-order filling-controlled MIT at $T>0$ we find hysteresis,
  i.e., the densities $n_c^{met}$ and $ n_c^{iso}$ in the two phases
  at the transition $\mu_c$ are different.  The same behavior has been
  obtained in quantum Monte Carlo (QMC) DMFT
  calculations,\cite{kotliar-2002,werner-2006} and with the dynamical
  cluster approximation (DCA) for the two-dimensional Hubbard
  model.\cite{macridin-2006} In the latter work the similarity
  of the $n(\mu)$ curves to the $p$-$V$ isotherms of a van-der-Waals
  gas was noted.  The insulating solution is characterized by a low
  charge compressibility and thus corresponds to the (incompressible)
  liquid, while in the metal $\chi_{\mu}$ stays finite at the
  transition even for $T\to 0$.

\begin{figure}[t]
    \centerline{\includegraphics[clip,width=\picturewidth,angle=0]{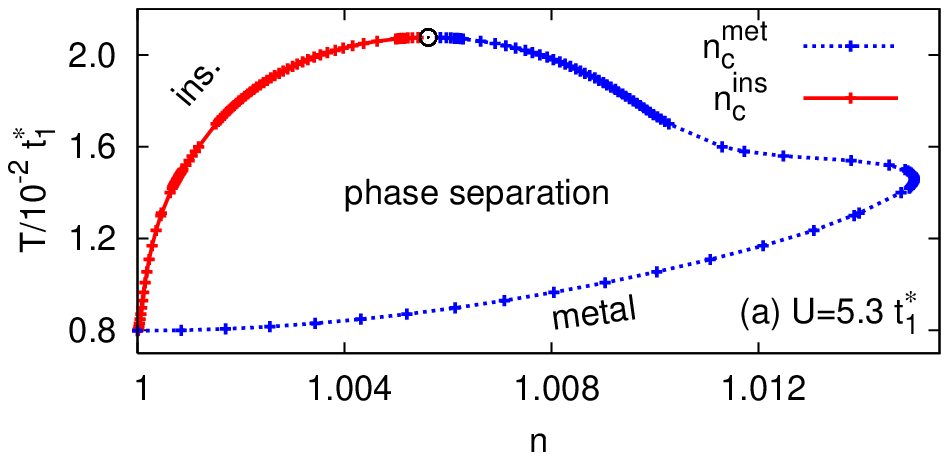}}
    \centerline{\includegraphics[clip,width=\picturewidth,angle=0]{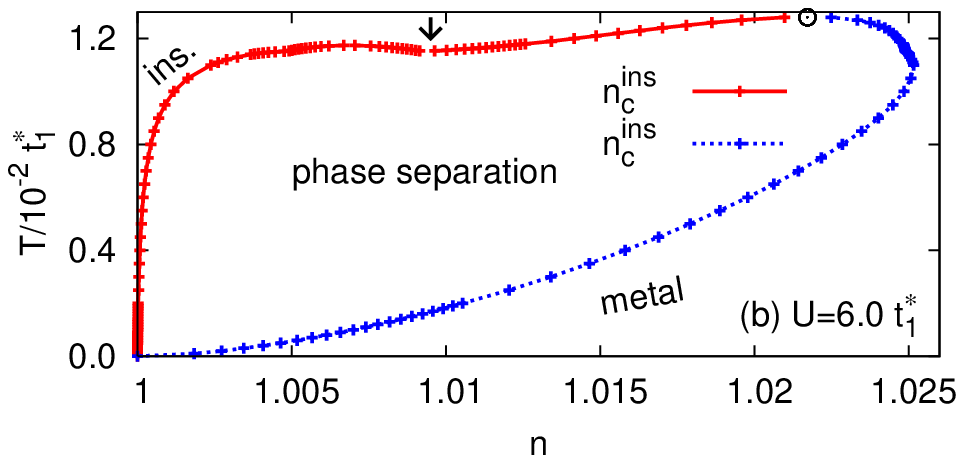}}
    \caption{($T,n$) phase diagram for $U=5.3\,t_1^*$ and
      $U=6\,t_1^*$; both for $t_2=0$.  Small circles indicate the
      critical point, corresponding to the second-order endpoint of
      the transition line in the ($T,\mu$) phase diagram.  The small
      dip in $n_c^{ins}$ for $U=6\,t_1^*$ (arrow) corresponds to an
      intermediate phase and might be an artefact of the
      two-site approximation. }
    \label{fig:nn}
  \end{figure}

  If the system is prepared at a density within the discontinuities
  [$n_c^{ins},n_c^{met}$], the free energy is minimized by the
  formation of a phase mixture with a fraction $x_{met} =
  (n-n_c^{ins})/(n_c^{met}-n_c^{ins})$ of the metal and $1-x_{met}$ of
  the insulator.  With QMC it is difficult to go beyond a calculation
  of $n(\mu)$ and determine also $\mu_c$, $n_c^{met}$, and
  $n_c^{ins}$. Within the SFA the latter quantities can easily be
  calculated, since this requires only a comparison of the grand
  potential (per site) $\Omega$ in the two phases, a quantity that is 
  calculated with high precision anyway.  This allows us to go from
  the ($T$,$\mu$) phase diagram to the ($T$,$n$) phase diagram, which
  displays the region between $n_c^{met}(T)$ and $n_c^{ins}(T)$ where
  the system is unstable against phase separation (Fig.~\ref{fig:nn}).
  Note that phase separation occurs only away from half-filling.  This
  will be discussed further in the next section.

  \section{NN and NNN hopping}
  \label{section:t1t2}

  In the following section we investigate the influence of NNN hopping
  on the paramagnetic phase diagram.  We focus on the case of
  $t_2^*/t_1^*=3/7$. Other values of $t_2$ yield qualitatively similar 
  results.  The energy scale is set to the square root of
  the second moment\cite{bluemer-phd} of the noninteracting DOS,
  $t^*\equiv\sqrt{t_1^{*2}+t_2^{*2}}$.
  \begin{figure}[t]
    \centerline{\includegraphics[clip,width=\picturewidth,angle=0]{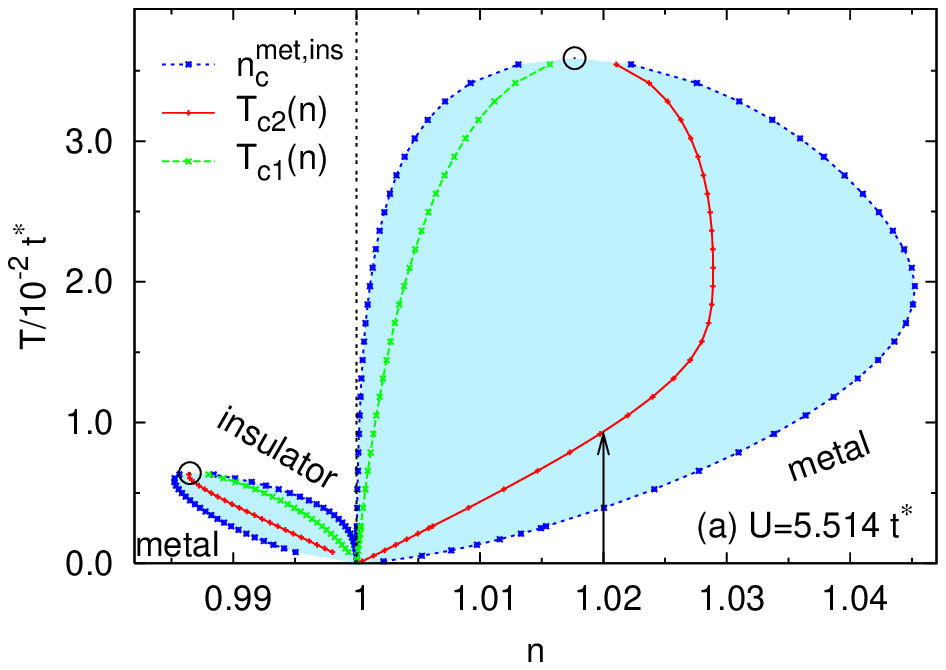}}
    \centerline{\includegraphics[clip,width=\picturewidth,angle=0]{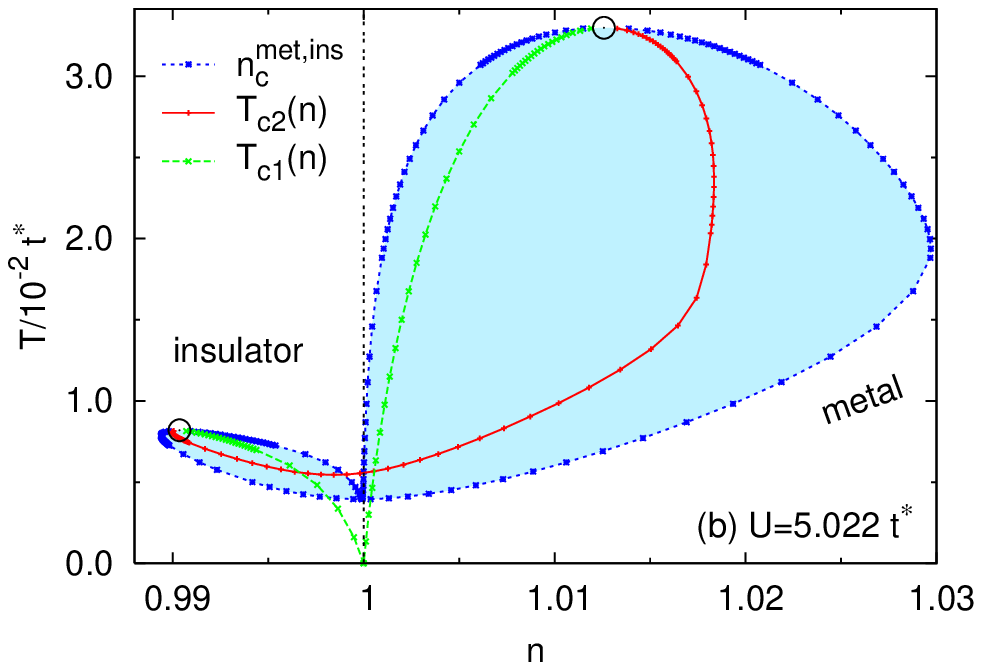}}
    \caption{($T,n$) phase diagram for $t_2^*/t_1^*=3/7$.  (a)
      $U=5.514\,t^*$ and (b) $U=5.022\,t^*$. The dotted line is the
      thermodynamic transition line ($n_c^{met}$ and $n_c^{ins}$).
      Inside this line (shaded region) the system is unstable against
      phase separation.  The arrow in (a) is referred to in the text.}
    \label{fig:t2a}
  \end{figure}

  \subsection{Filling-controlled transition}
  \label{subsection:NNN-phase-diagram}

  For $t_2\ne 0$ the surface $T_c(U,\mu)$ of the first-order
  transition in the ($U,T,\mu$) phase diagram retains the overall
  shape as in Fig.~\ref{fig:NN-UTmu}, but is no longer symmetric with
  respect to $\mu=U/2$ due to particle-hole asymmetry.  The value of
  the critical interaction $U_{c2}$ turns out to be rather insensitive
  to the value of $t_2$: for $t_2^*/t_1^*=3/7$ we find $U_{c}\approx
  5.45 t^*$, in comparison to $U_{c}\approx 5.88 t_1^*$ for pure NN
  hopping.  Also the ($T,n$) phase diagrams look very similar to that
  for of pure NN hopping (Fig.~\ref{fig:t2a}).  However, particle-hole
  asymmetry is immediately evident by the very different sizes of the
  two phase-separated regions, which enclose the Mott transition
  driven by hole- and electron-doping, respectively.

  In addition to the phase-separated region, also the spinodal curves
  $T_{c1}(n)$ and $T_{c2}(n)$ are included in Fig.~\ref{fig:t2a}.  To
  clarify their meaning let us consider the behavior of the system
  upon heating at given density $n=n_0$, starting from the metallic
  phase at $T=0$ (arrow in Fig.~\ref{fig:t2a}a).  When
  $n_0=n_c^{met}(T)$ the system becomes thermodynamically unstable
  against the formation of a phase mixture containing the metallic
  phase ($n>n_0$) and the insulating phase ($n \approx 1$).  However,
  the purely metallic phase ($n=n_0$) can be superheated up to the
  spinodal curve $T_{c2}$.  The charge compressibility diverges at the
  spinodal curve, indicating increasing density fluctuations in the
  metastable metallic phase.  For $T>T_{c2}$ a purely metallic
  solution no longer exists.  Similarly, the insulating phase can be
  supercooled down to $T_{c1}$.
  \begin{figure}[t]
    \centerline{\includegraphics[clip,width=\picturewidth,angle=0]{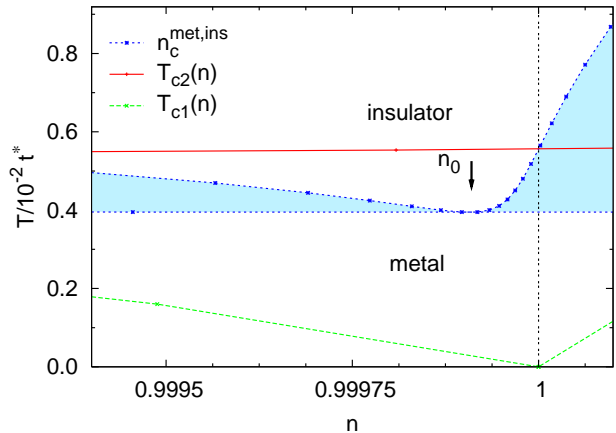}}
    \caption{Closeup of Fig.~\ref{fig:t2a}b near $n=1$,
      for a  discussion see text.}
    \label{fig:u3.825}
  \end{figure}

  For the smaller value of the two interactions in Fig.~\ref{fig:t2a},
  a closer look at the interval around $n=1$ reveals a novel situation
  compared to the case of $t_2=0$ (cf. Fig.~\ref{fig:u3.825}).
  Obviously, phase separation now occurs at half-filling, but
  disappears at a special density $n_0\ne1$.  To understand these
  numerical results it is convenient to go back to the ($T$,$\mu$)
  representation.  For given $U$ and $t_2=0$, the transition line
  $T_c(\mu)$ has a characteristic U-shape (Fig.~\ref{fig:NN-UTmu}),
  where the minimum is at $\mu=U/2$.  Precisely at this minimum no
  phase separation is observed. This is clear from the
  Clausius-Clapeyron equation
  \begin{equation}
    \frac{dT_c}{d\mu} = \frac{n_c^{met}-n_c^{ins}}{S_c^{ins}-S_c^{met}},
    \label{eq:clausius}
  \end{equation}
  which relates the slope of $T_{c}(\mu)$ to the density $n_c^{met}$
  ($n_c^{ins}$) and the entropy (per site) $S_c^{met}$ ($S_c^{ins}$) in the two
  phases at the transition.  The U-shape has a simple thermodynamic
  explanation: The stable phase is determined by its lower grand
  potential (per site) $\Omega = E - n\mu - TS$.  For $U<U_{c2}$ we know that the
  metal is stable at $T=0$.  Let now $\mu=\mu_1$ be such that
  $n^{met}=1$ for $T=0$.  We can expand the difference
  $\Delta\Omega(\mu)\equiv(\Omega^{met}-\Omega^{ins})_{T=0}$ as
  \begin{equation}
    \Delta\Omega(\mu) = 
    \rm{const.}-\frac{\chi^{met}_\mu}{2} (\mu-\mu_1)^2+\ldots,
    \label{eq:om0}
  \end{equation}
  where we used $\partial\Omega/\partial\mu=-n$ as well as
  $n^{ins}_{T=0}=1$.  The metal thus becomes more stable as $n$
  deviates from $n=1$.  However, in DMFT the insulator has a large
  entropy even at $T=0$ since there is no ordering of the magnetic
  moments.  This entropy gain stabilizes the insulator at a certain
  temperature $T>0$ and causes the first-order transition.  The
  minimum of $T_c$ directly reflects the minimum in
  $\Delta\Omega(\mu)$.  All this remains true for $t_2\ne0$.  However,
  while the minimum of $T_c$ lies at $n=1$ as a consequence of
  particle-hole symmetry for $t_2=0$, it shifts to $n_0\ne1$ for
  $t_2\ne0$.

  \subsection{Half-filling ($n=1$)}
  \label{subsection:NNN-half-filling}

  Phase separation makes the calculation of the phase diagram for a
  fixed density considerably more difficult.  It is impossible to
  decide if a system at some given density prefers to assume an
  inhomogeneous state by investigating only that particular density.
  Instead, we construct the ($T$,$U$) phase diagram for half-filling
  from several ($T$,$n$) phase diagrams (for various $U$) as in
  Fig.~\ref{fig:t2a}.
  \begin{figure}[t]
    \centerline{\includegraphics[clip=,width=\picturewidth,angle=0,bbllx=68,bblly=60,bburx=405,bbury=294]{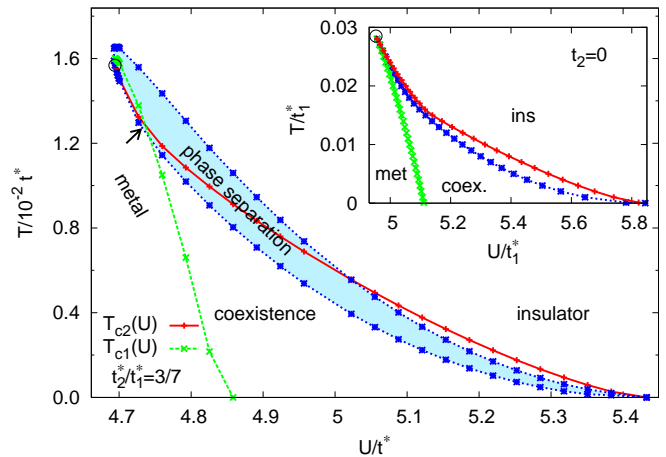}}
    \caption{($T,U$) phase diagram for half filling and $t_2^*/t_1^*=3/7$.
      For comparison, the corresponding plot for pure NN hopping
      obtained within the two-site approximation is shown
      in the inset. The circle denotes the
      critical point. Note that the spinodal curves cross
      once within the phase separated region (arrow).}
    \label{fig:n=1}
  \end{figure}
  The first-order line of the particle-hole symmetric case is then
  found to be replaced by a phase-separated region as shown in
  Fig.~\ref{fig:n=1}.  Within this region, the stable phase is a
  mixture of two slightly-doped phases. Their composition cannot be
  inferred from Fig.~\ref{fig:n=1} alone and again requires ($T$,$n$)
  sections.  A distinctive feature of the phase diagram for pure NN
  hopping at half-filling is the triangular-shaped coexistence region
  (see Fig.~\ref{fig:NN-UTmu}).  By contrast, in the presence of
  frustration we find regions of phase coexistence, as well as a small
  region (close to the critical point) in which both metal and
  insulator phase cannot exist as pure phases at $n=1$. This is
  because the spinodal lines cross once and meet at the critical
  point, which is indicated by a circle in Fig~\ref{fig:n=1}.

\begin{figure}[t]
    \centerline{\includegraphics[clip,width=\picturewidth,angle=0]{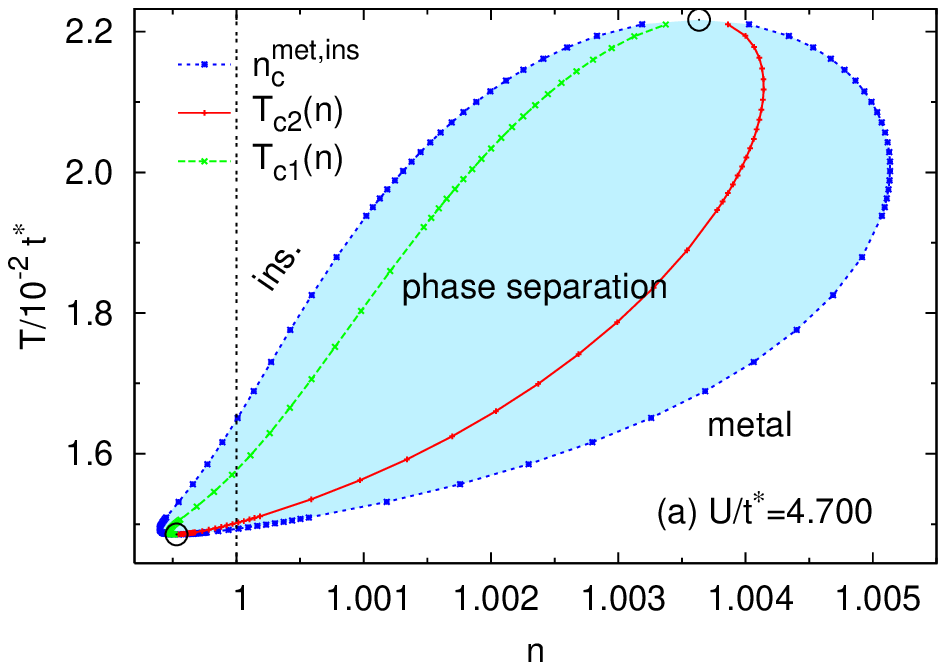}}
    \centerline{\includegraphics[clip,width=\picturewidth,angle=0]{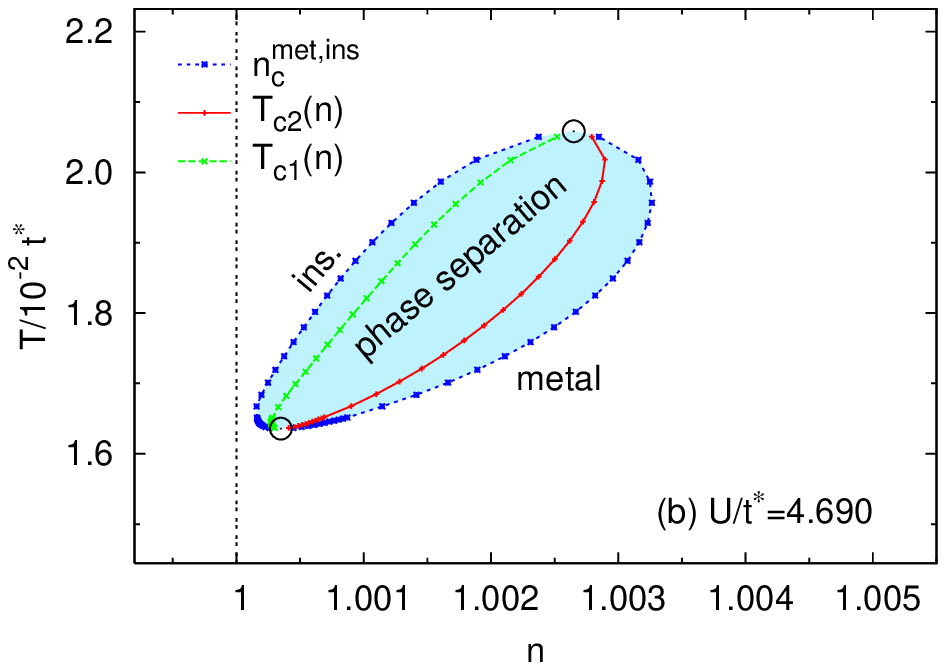}}
    \caption{($T$,$n$) phase diagram for $t_2^*/t_1^*=3/7$. (a)
      $U=4.7\,t^*$ and (b) $U=4.69\,t^*$.}
    \label{fig:t2b}
  \end{figure}

  For a closer look at the critical region consider
  Fig.~\ref{fig:t2b}.  When $U$ is lowered the transition shifts to
  larger temperatures and becomes more and more asymmetric
  (Fig.~\ref{fig:t2b}a), until it finally occurs only in the
  electron-doped regime (Fig.~\ref{fig:t2b}b).  For a given $U$, the
  first-order line $T_{c}(\mu)$ has two critical endpoints. In general
  the density at these points, i.e., the critical density
  $n_{crit}(U)$ is different from one.  However, when $n_{crit}=1$
  (which occurs between $U=4.69\,t^*$ and $U=4.7\,t^*$ in the present
  case), there is a critical point in the ($T$,$U$) phase diagram for
  half-filling.  This point is then characterized by a diverging
  charge compressibility $\chi_\mu$.

  Quantitatively, NNN hopping leads to a considerable suppression of
  the transition region to lower temperatures, while the critical
  interaction does not change very much.  How much this will be
  modified when one goes beyond the two-site approximation remains to
  be investigated.  Nevertheless, the fact that NNN hopping reduces the
  Mott transition temperature implies that the competition between
  paramagnetic and antiferromagnetic phases for $t_1$-$t_2$ hopping is
  more complicated than previously expected.

  \section{Conclusion}
  \label{section:conclusion}

  We employed the self-energy functional approach in combination with
  dynamical mean-field theory to compute the metal-insulator
  transition in the ($U$,$T$,$n$) phase diagram of correlated
  electrons. To this end the single-band Hubbard model with NN and NNN
  hopping on a Bethe lattice was investigated in the paramagnetic
  phase.  For pure NN hopping we find a transition scenario that is
  consistent with previous calculations, namely a first-order
  filling-controlled metal-insulator transition with phase separation.
  No phase separation occurs at half-filling, where the transition as
  a function of the interaction strength is also first order for finite temperatures.
  NNN hopping strongly modifies this picture and leads to phase separation
  already at half-filling, i.e., to a discontinuity in $n(\mu)$ such
  that $n(\mu-\eta)<1<n(\mu+\eta)$, $\eta\to0^+$.  Close to the
  critical point the phase diagram for half-filling and $t_2\ne0$
  resembles that for $t_2=0$ and low doping.  We note that for the
  Hubbard model with pure NN hopping in $d=2$ phase separation was
  excluded analytically at half-filling,\cite{su-1996} but this
  argument does not apply to $d\ge3$.

  It is clear that in reality the long-range nature of the Coulomb
  interaction will counteract the formation of a strongly
  inhomogeneous charge distribution.  Nevertheless the instability
  with respect to phase separation found in the Hubbard model hints at
  intriguing physics.  Namely, it shows that the inclusion of further
  interactions will lead to new and interesting phenomena, e.g., more
  complex ordering phenomena.

  Another open question is the influence of NNN hopping on the
  antiferromagnetic phase.  QMC results indicate the suppression of
  antiferromagnetism.\cite{radke-phd,schlipf-phd} However, to decide
  whether this will be strong enough to reveal the metal-insulator
  transition in the paramagnetic phase, the SFA needs to be evaluated
  for larger reference systems.  The present results suggest
  that frustration not only suppresses the temperature of the 
  antiferromagnetic transition but also that of the Mott transition.
  Apparently NNN hopping leads to a more subtle competition between
  metallic, insulating, and magnetically ordered phases than
  previously thought. Therefore it remains a challenging task to
  understand the physical properties of strongly correlated materials
  such as $\rm V_2O_3$ in terms of a minimal electronic correlation
  model.

  \section*{Acknowledgements}

  We are grateful to Ralf Bulla and Krzysztof Byczuk for discussions.
  The work was supported in part by Deutsche Forschungsgemeinschaft
  (DFG) through Sonderforschungsbereich 484.


\begin{thebibliography}{99}

  \bibitem{mott}
    N.\ Mott,
    {\em Metal-Insulator Transitions},
    Taylor and Francis, London, 1990.

  \bibitem{imada-1998}
    M.\ Imada, A.\ Fujimori, and Y.\ Tokura,
    Rev.\ Mod.\ Phys.\ {\bf 70}, 1039 (1998).

  \bibitem{hubbard-1963}
    J.\ Hubbard,
    Proc.\ Roy.\ Soc.\ London Ser.\ A {\bf 276}, 238 (1963).

  \bibitem{metzner-1989b}
    W.\ Metzner and D.\ Vollhardt,
    Phys.\ Rev.\ Lett.\ {\bf 62}, 324 (1989).

  \bibitem{georges-1996}
    A.\ Georges, G.\ Kotliar, W.\ Krauth, and M.~J.\ Rozenberg,
    Rev.\ Mod.\ Phys.\ {\bf 68}, 13 (1996).

  \bibitem{kotliar-2004}
    G.\ Kotliar and D.\ Vollhardt,
    Phys.\ Today {\bf 57}, 53 (2004).

  \bibitem{jarrell-1992}
    M.\ Jarrell,
    Phys.\ Rev.\ Lett.\ {\bf 69}, 168 (1992).
    
  \bibitem{georges-1993}
    A.\ Georges, and W.\ Krauth,
    Phys.\ Rev.\ B {\bf 48}, 7167 (1993).
    
  \bibitem{rozenberg-1994}
    M.~J.\ Rozenberg, G.\ Kotliar, and X.~Y.\ Zhang,
    Phys.\ Rev.\ B {\bf 49}, 10181 (1994).

  \bibitem{bulla-1999}
    R.\ Bulla,
    Phys.\ Rev.\ Lett {\bf 83}, 136 (1999).

  \bibitem{rozenberg-1999}
    M.~J.\ Rozenberg, R.\ Chitra, and G.\ Kotliar,
    Phys.\ Rev.\ Lett {\bf 83}, 3498 (1999).

  \bibitem{bulla-2001}
    R.\ Bulla, T.~A.\ Costi, and D.\ Vollhardt,
    Phys.\ Rev.\ B {\bf 64}, 045103 (2001).

  \bibitem{tong-2001}
    N.-H.\ Tong, S.-Q.\ Shen, and F.-C.\ Pu,
    Phys.\ Rev.\ B {\bf 64}, 235109 (2001).

  \bibitem{joo-2001}
    J.\ Joo and V.\ Oudovenko,
    Phys.\ Rev.\ B {\bf 64}, 193102 (2001).

  \bibitem{bluemer-phd}
    N.\ Bl\"{u}mer,
    Ph.~D.\ thesis, Universit{\"a}t Augsburg, 2002.

  \bibitem{fisher-1995}
    D.~S.\ Fisher, G.\ Kotliar, and G.\ Moeller,
    Phys.\ Rev.\ B {\bf 52}, 17112 (1995).

  \bibitem{kajueter-1996}
    H.\ Kajueter, G.\ Kotliar, and G.\ Moeller,
    Phys.\ Rev.\ B {\bf 53}, 16214 (1996).

  \bibitem{ono-2001}
    Y.\ Ono, R.\ Bulla, A.\ Hewson, and M.\ Potthoff,
    Eur.\ Phys.\ J.\ B {\bf 22}, 283 (2001).

  \bibitem{kotliar-2002}
    G.\ Kotliar, S.\ Murthy, and M.~J.\ Rozenberg,
    Phys.\ Rev.\ Lett.\ {\bf 89}, 046401 (2002).
 
  \bibitem{garcia-2006}
    D.~J.\ Garc\'{i}a, E.\ Miranda, K.\ Hallberg, and M.~J.\ Rozenberg,
    cond-mat/0608248.
  
  \bibitem{werner-2006}
    P.\ Werner and A.~J.\ Millis,
    cond-mat/0610401.
    
  \bibitem{ohashi-2006}
    T.\ Ohashi and N.\ Kawakami,
    Phys.\ Rev.\ Lett.\ {\bf 97}, 066401 (2006).

  \bibitem{aryanpour-2006}
    K.\ Aryanpour, W.~E.\ Pickett, and R.~T.\ Scalettar,
    cond-mat/0604609.

  \bibitem{rozenberg-1995}
    M.~J.\ Rozenberg, G.\ Kotliar, H.\ Kajueter,
    G.~A.\ Thomas, D.~H.\ Rapkine, J.~M.\ Honig, and P.\ Metcalf,
    Phys.\ Rev.\ Lett.\ {\bf 75}, 105 (1995).

  \bibitem{hirsch-2002}
    J.~E.\ Hirsch,
    Phys.\ Rev.\ B {\bf 65}, 184502 (2002).

  \bibitem{hirsch-2005}
    J.~E.\ Hirsch,
    Phys.\ Rev.\ B {\bf 71}, 104522 (2005).

  \bibitem{chitra-1999}
    R.\ Chitra and G.\ Kotliar,
    Phys.\ Rev.\ Lett.\ {\bf 83}, 2386 (1999).

  \bibitem{zitzler-2004}
    R.\ Zitzler, N.~H.\ Tong, T.\ Pruschke, and R.\ Bulla,
    Phys.\ Rev.\ Lett.\ {\bf 93}, 016406 (2004).

  \bibitem{eckstein-2005}
    M.\ Eckstein, M.\ Kollar, K.\ Byczuk, and D.\ Vollhardt,
    Phys.\ Rev.\ B {\bf 71}, 235119 (2005).

  \bibitem{kollar-2005}
    M.\ Kollar, M.\ Eckstein, K.\ Byczuk, N.\ Bl{\"u}mer,   
    P.\ van~Dongen, M.~H.\ Radke~de~Cuba, W.\ Metzner,   
    D.\ Tanaskovic, V.\ Dobrosavljevic, G.\ Kotliar, and D.\ Vollhardt,
    Ann.\ Phys.\ {\bf 14}, 642 (2005).

  \bibitem{macridin-2006}
    A.\ Macridin, M.\ Jarrell, and T.\ Maier,
    Phys.\ Rev.\ B {\bf 74}, 085104 (2006).

  \bibitem{moreo-1991}
    A.\ Moreo, D.\ Scalapino, and E.\ Dagotto,
    Phys.\ Rev.\ B {\bf 43}, 11442 (1991).

  \bibitem{becca-2000}
    F.\ Becca, M.\ Capone, and S.\ Sorella,
    Phys.\ Rev.\ B {\bf 62}, 12700 (2000).

  \bibitem{potthoff-2003a}
    M.\ Potthoff,
    Eur.\ Phys.\ J.\ B {\bf 32}, 429 (2003).

  \bibitem{luttinger-1960}
    J.~M.\ Luttinger and J.~C.\ Ward,
    Phys.\ Rev.\ {\bf 118}, 1417 (1960).

  \bibitem{potthoff-2003c}
    M.\ Potthoff, M.\ Aichhorn, and C.\ Dahnken,
    Phys.\ Rev.\ Lett.\ {\bf 91}, 206402 (2003).

  \bibitem{tong-2005}
    N.~H.\ Tong,
    Phys.\ Rev.\ B {\bf 72}, 115104 (2005).

  \bibitem{potthoff-2003b}
    M.\ Potthoff,
    Eur.\ Phys.\ J.\ B {\bf 36}, 335 (2003).

  \bibitem{pozgajcic-2004}
    K.\ Po\v{z}gaj\v{c}i\'{c},
    cond-mat/0407172.

  \bibitem{potthoff-2004}
    M.\ Potthoff,
    Condens.\ Mat.\ Phys.\ {\bf 9}, 557 (2006).

  \bibitem{su-1996}
    G.\ Su,
    Phys.\ Rev.\ B {\bf 54}, R8281 (1996).

  \bibitem{radke-phd}
    M.~H.\ Radke~de~Cuba,
    Ph.~D.\ thesis, RWTH Aachen, 2002.

  \bibitem{schlipf-phd}
    J.-M.\ Schlipf,
    Ph.~D.\ thesis, Universit{\"a}t Augsburg, 1998.

  \end{thebibliography}
\end{document}